\documentclass[sigconf,screen]{acmart}
\usepackage{color, colortbl}
\usepackage{graphicx} % Required for inserting images
\usepackage{dblfloatfix} 
\usepackage{wrapfig} 
\usepackage{balance} 
\usepackage[utf8]{inputenc}
\usepackage{microtype}
\usepackage[normalem]{ulem}
\usepackage[T1]{fontenc}  
 
\setcopyright{acmlicensed}
\acmPrice{15.00}
\acmDOI{10.1145/3617555.3617876}
\acmYear{2023}
\copyrightyear{2023}
\acmSubmissionID{fsews23promisemain-p14-p}
\acmISBN{979-8-4007-0375-1/23/12}
\acmConference[PROMISE '23]{Proceedings of the 19th International Conference on Predictive Models and Data Analytics in Software Engineering}{December 8, 2023}{San Francisco, CA, USA}
\acmBooktitle{Proceedings of the 19th International Conference on Predictive Models and Data Analytics in Software Engineering (PROMISE '23), December 8, 2023, San Francisco, CA, USA}
\received{2023-07-07}
\received[accepted]{2023-07-28}

\begin{abstract}
 To make models more understandable and correctable,   I propose that the PROMISE community pivots to the problem of {\em model review}. 
Over the years, there have been many reports that very simple models can perform exceptionally well. Yet, where are the researchers asking ``say, does
that mean that we could make software analytics simpler and more comprehensible?''
 This is an important question, since humans often have difficulty accurately assessing complex models (leading to unreliable and sometimes dangerous results). 
 
 Prior PROMISE results have shown that data mining can effectively summarizing large models/ data sets into simpler and smaller ones.  Therefore, the PROMISE community has the skills and experience needed to redefine, simplify, and improve the relationship between humans and AI.
\end{abstract}
 

 
\begin{document}
\title{Model Review: A PROMISEing Opportunity}

\author{Tim Menzies}
\orcid{0000-0002-5040-3196}
\email{timm@ieee.org}
\affiliation{%
  \institution{North Carolina State University}  
  \country{USA}
}
\maketitle
\pagestyle{plain} 
\section{Introduction}
PROMISE will soon enter its third decade. 
 What have we learned from two decades of PROMISE v1.0 that could shape the next decade of PROMISE v2.0?
 
Over the years, there have been many results
where   very simple models performed exceptionally well~\cite{Holte1993VerySC,menzies2008implications,agrawal2019dodge,Xu21,Tawosi23,Kohavi97}.
Using those results,   PROMISE v2.0 could focus less on model creation,  and more on something that requires and demands simpler and more comprehensible models. 
Specifically, I say PROMISE v2.0 should be about simplifying  {\em model review} (using data mining).

% It may seem disingenuous, even dishonest, when someone warns of a problem that they themselves have caused. For example, President Eisenhower's 1961 farewell address warned on the dangers of the military-industrial complex\footnote{Eisenhower said ``We must guard against the acquisition of unwarranted influence, whether sought or unsought, by the military-industrial complex. The potential for the disastrous rise of misplaced power exists and will persist.''}-- an institution that was partially brought about by his administration's budget priorities.  

% I say this since the theme of this paper is that there are problems with the PROMISE conference series; and that I am (partially)
% responsible for those problems.  Mea culpa.
% But the good news is that, in my view, these problems can be fixed using tools developed from the PROMISE experience.

\section{  Why Change   PROMISE?}

  PROMISE v1.0 was created one night in 2004 walking around Chicago's Grant Park. Jelber Sayyad-Shirabad and I had spent the day at a disappointing workshop
on data mining and SE. ``Must do better'', we said.  ``Why don't we do it  like in ML? Make conclusions reproducible? Demand that if people publish a paper, they should also publish the data used in that data?"\footnote{
In 2023 it is  hard to believe that ``reproducible SE'' was a radical idea.  But once upon a time, there was little sharing of data and scripts-- so much so that in 2006 Lionel Briand predicted the failure of PROMISE saying ``no one will give you data''.}.

At first, the series got off to a shaky  start. But once
Elaine Weyuker, Thomas Ostrand, Gary  Boetticher, and Guenther Ruhe
joined the steering committee, the meeting earned the prestige needed for future  growth.
And
in those early days, it
was impressive to see so many researchers
taking up the idea of reproducible results.
Numerous papers were written that applied an increasing
elaborate tool set to data like COC81, JM1, XALAN, DESHARNIS
and all the other data sets that were used (and reused) in the first decade of PROMISE.

Many of those papers lead to successful research. In 2018, 20\% of the articles listed in {\em Google Scholar Software Metrics} for {\em IEEE Transactions
on SE} used data sets from the first decade of PROMISE. So while other research areas struggled to obtain reproducible results, PROMISE swam (as it were) in an ocean of reproducibility.

The problem was that in the second decade of PROMISE, many researchers still continue that kind of first-decade research. For example,   all too often, I must review papers from authors who think it is valid
to publish results based on (e.g.) the COC81 data set first published in 1981~\cite{Boehm1981}; the DESHARNIS data set,  first published in 1989~\cite{desharnais1989analyse}; the JM1 data, first published in 2004~\cite{menzies2004good}; or the XALAN data set, first published in 2010~\cite{jureczko2010towards}\footnote{Just to be clear,
there is value in   a publicly accessible collection of reference problems. For instance, if a PROMISE author is unable to present results from confidential industrial data, they can use the reference collection to construct a reproducible example of their technique. That said, I am usually tempted to reject papers that are {\em solely} based on defect datasets that I contributed to the PROMISE repository in 2005 (e.g. CM1, JM1, KC1, KC2, KC3, KC4, MC1, MC2, MW1, PC1, PC2, PC3, PC4 and PC5)
since, in 2023 we have access to much more recent data
(e.g. see the 1100+ recent Github projects
used by Xia et al.~\cite{xia22}).}.

% Another conclusion from the first decade of PROMISE was that we might have been exploring the wrong problem. As noted in an editorial on the best papers from PROMISE 2011~\cite{10.1016/j.infsof.2013.03.006}, the initial belief was that as we collected more data, PROMISE researchers could decrease the variance in their conclusions. However, the opposite turned out to be the case: the {\em more} data we collected,
% the {\em more diverse} that data became and the {\em larger} were the measured variances. I
% tried explaining this, saying   SE learning should focus on  
%  local clusters and not  global models~\cite{menzies2011local,menzies2012local}. Others agreed ~\cite{10.1145/2372251.2372256,bettenburg2012think} while others found no evidence that local models generated better
% predictions than global models~\cite{herbold2017global}.   

% @inproceedings{canfora2013multi,
%   title={Multi-objective cross-project defect prediction},
%   author={Canfora, Gerardo and De Lucia, Andrea and Di Penta, Massimiliano and Oliveto, Rocco and Panichella, Annibale and Panichella, Sebastiano},
%   booktitle={2013 IEEE Sixth International Conference on Software Testing, Verification and Validation},
%   pages={252--261},
%   year={2013},
%   organization={IEEE}
% }
 
Meanwhile, AI fever took over SE. As of 2018,
it became standard at ASE, FSE, ICSME, ICSE, etc. to see papers
that make much use of AI. For example, the MSR conference (which in the early days looked like a sister event to PROMISE) has grown to a large annual A-grade venue.
And just as MSR grew, so did PROMISE shrink. In 2008, PROMISE was a two-day event with 70 attendees. PROMISE is now a much smaller and shorter event. Without definitive results or a novel technological position, it became difficult to differentiate PROMISE from dozens of other, somewhat more prominent, venues.

\section{  Why Switch to  Model Review?}

Reducing the size of the model is an important part of model review.
According to psychological theory~\cite{czerlinski1999good, gigerenzer1999good, martignon2003naive, brighton2006robust, martignon2008categorization, gigerenzer2008heuristics, phillips2017FFTrees, gigerenzer2011heuristic,neth2015heuristics},
humans can best review a system when it ``fits'' it into their memory; i.e., when that it comprises many small model
fragments. 
 Larkin et al.~\cite{Larkin1335} characterize human expertise by a very large long-term memory (LTM)
and a very small short-term memory (STM) that contains as few as four to seven items\footnote{Ma et al.~\cite{ma2014changing} 
 have used evidence from neuroscience and functional MRIs to
argue that the capacity of the STM could be better measured using other factors than ``number of
items ``''. But even they conceded that ``the concept of limited (STM) has considerable
explanatory power for behavioral data''.}.
The LTM contains  many tiny rule fragments
that explore the contents
of the STM to say ``when you see THIS, do THAT''.
When an LTM rule triggers,  can rewrite the STM content, which,
in turn, can trigger other rules.
Experts are experts, says Larkin et al.~\cite{Larkin1335} because LTM patterns dictate what to do, without having to pause for reflection. Novices perform worse than experts,
says Larkin et al., when they fill up their STM with too many to-do's where they plan to pause and reflect on what to do next.  
This theory is widely endorsed.  
 Phillips et al.~\cite{phillips2017FFTrees} discuss how models with tiny fragments can be quickly comprehended by emergency room physicians making rapid decisions; or by guard soldiers making snap decisions about whether to fire or not on a potential enemy; or by stockbrokers making instant decisions about buying or selling stock. 

Complex models cannot fit into STM, leading   to problems with model review.
Green~\cite{green2022flaws}  comments that when faced with large and complex problems,
  cognitive theory~\cite{simon1956rational} tells us that humans use heuristic ``cues'' to lead them to the most important parts
of a model.   Such cues are essential if humans are to reason about large problems.   That said,  using cues can introduce their own errors:
   {\em 
   ...people (including experts) are susceptible to ``automation bias'' (involving) omission errors - failing to take action because the automated system did not provide an alert - and commission error}~\cite{green2022flaws}.
 This means that oversight can lead to the opposite desired effect by {\em ``legitimizing the use of faulty and controversial (models) without addressing (their fundamental issues'')}~\cite{green2022flaws}. 

By ``faulty and controversial models'', Green refers to
the  long list of examples where detrimental models  were learned via
algorithmic means.
For example,   Cruz et al.~\cite{cruz2021promoting}  lists examples where:
\begin{itemize}
\item Proposals from low-income groups are   are five times more  likely     to be incorrectly ignored by donation groups;
\item Woman can be five times more likely to be incorrectly classified as low income;
\item African Americans are five times more likely to  languish in prison until trial,
rather given the bail they deserve.  
\end{itemize}
These are just a few of the many reported examples\cite{rudin2019explaining}\footnote{See also 
\url{http://tiny.cc/bad23a}, \url{http://tiny.cc/bad23b}, \url{http://tiny.cc/bad23c}} of algorithmic discrimination.
For another example, the last chapter of Noble~\cite{noble2018algorithms} describes how a successful
hair salon went bankrupt due to internal
choices within the YELP recommendation algorithm.

Like
Mathews~\cite{mathews23}, I am not surprised that so many models are unfair. 
Mathews argues that everyone seeks ways to exploit some advantage for themselves. Hence, we should
expect that the software we build to discriminate against some social groupings;
\begin{quote}

{\em 
    ``People often think of their own hard work or a good decision they made. However, it is often more accurate to look at advantages like the ability to borrow money from family and friends when you are in trouble, deep network connections so that you hear about opportunities or have a human look at your application,  the ability to move on from a mistake that might send someone else to jail, help at home to care for children,  etc. The narrative that success comes from hard work misses that many people work hard and never succeed. Success often comes from exploiting a playing field that is far from level and when push comes to shove, we often want those advantages for our children, our family, our friends, our community, our organizations.''}
 \end{quote}
Hence I assert that
unfairness is a widespread issue that needs to be addressed and managed. 
Specifically, we need 
to ensure that a software system created by one group, $A$, can be critiqued and modified by another group, $B$.
There are many ways to do this and at PROMISE'23, it is appropriate to focus on the data mining methods issues
(see next section). 

But first, we take care to stress that technical methods like data mining should not be used in isolation.
 PROMISE v2.0 should acknowledge its
   relationship and   responsibilities to those affected by the tools we deliver. We must stop ``flattening'', which is the trivialization and even ignoring of legitimate complaints that our institutions discriminate against certain social groups~\cite{Coaston19}. Bowleg~\cite{doi:10.2105/AJPH.2020.306031} warns that flattening ``depoliticized and stripped (its) attention to power, social justice, and (how we do things wrongly)''.
   
To address these issues, organizations should review their hiring practices to diversify the range of perspectives seen in design teams.  Requirements engineering practices should be improved to include extensive communication with the stakeholders of the software. 
 Software testing teams should extend their tests to cover issues such as discrimination against specific social groups~\cite{cruz2021promoting,10.1145/3585006,Chakraborty}. 
 
Further, on the legal front,
Canellas~\cite{canellas21} and Mathews et al.~\cite{matthewsshould} suggest a tiered process in which the most potentially discriminatory projects are routinely reviewed by an independent external review team
(as done in the IEEE 1012 independent V\&V standard). Ben Green~\cite{green2022flaws} notes that reviewing software systems and AI systems is becoming a legislative necessity and that human-in-the-loop auditing of decisions made by software is often mandatory.
Such legislation is necessary to move away from the internal application of voluntary industrial standards (since, as seen in the Volkswagen emissions scandal (see \url{http://tiny.cc/scandalvw}), companies cannot always be trusted
to voluntarily apply reasonable standards.

\section{How To Make Smaller Models?}

How to make models smaller, more comprehensible, and easier to review?
 Enter all the data mining tools explored at PROMISE.
In those explorations, often it was seen that

 \centerline{\bf A small number of \underline{key} variables control the rest.}
 \noindent
Just to state the obvious, there is a clear
connection between  keys and the  cognitive effort needed  better review methods. Specifically: for a system
with keys, we only need to look
at a few keys to 
  understanding and/or critique and/or control  that system. 

Outside of SE, I have seen keys in hospital nutrition analysis~\cite{partington2015reduced} and avionic control systems~\cite{gay2010automatically}. Within SE, I have seen keys in:
\begin{itemize}
 \item  
Defect prediction datasets~\cite{menzies2006data} where two to three attributes were
enough to predict for defects
\item
Effort estimation models~\cite{chen2005finding} where four to eight attributes
where enough to predict for defects;
\item 
Requirements models for NASA deep space missions~\cite{jalali2008optimizing} where
two-thirds of the decision attributes could be ignored while still finding effective 
optimizations; 
\item
11 Github issue close time data sets~\cite{rees2017better}  
where only 3 attributes (median) were needed for   effective prediction.
\end{itemize}
One way to see how many keys are in a system is to ask how many {\em prototypes} (minimum number of examples)  are required to explore that system. 
At PROMISE"08, the keys were found in 50
examples (selected at random) since the models built from this small sample performed no worse than the models learned from thousands of additional examples~\cite{menzies2008implications}.  
In 2023 we made a similar observation. 
In a study that explored 20 years of data from 250 Github projects with 6000 commits per project (average).
In that study, the defect models learned from only the first 150 commits were predicted, as well as the models learned from much larger samples~\cite{10.1145/3583565}.

\begin{table}[!t]
\begin{tabular}{|p{.98\linewidth}|}\hline
\rowcolor{blue!10}
My preferred 
recursive bi-clustering procedure is  based on the FASTMAP Nystr{\"o}m algorithm~\cite{faloutsos1995fastmap,platt2005fastmap,papakroni2013data}. \\
This method maps points to the dimension
of greatest variance, computed as follows. Find any point.
Finds its furthest neighbor $A$, then finds $A$'s furthest neighbor
$B$. Map all other   points $X$   to the line  $\overline{AB}$ using 
$x=(a^2+c^2-b^2)/(2c)$ where $a,b,c$=${\mathit dist}(X,A), {\mathit dist}(X,B), {\mathit dist}(A,B)$. 
After mapping, we divide the data on the median $x$ value.
 \\
 \rowcolor{blue!10} The generation of contrast rules can then be used to report the minimal differences that
most distinguish the  {\em desired} and {\em current} leaf clusters. Specifically,
treat two leaf clusters {\em current} and the {\em desired} as a two-class
system. Using data just from those two clusters,
apply supervised discretization based on entropy to convert all attributes
into ranges. For ranges that appear
at frequency $x>0$ in the {\em desired}, rank them according to their frequency
$y$ in the {\em current}   using $x^2/(x+y)$.
  Build rules by exploring
all combinations of the (say) $N=10$ top-ranked ranges. 
\\
Pruning heuristics can be used to report only the essential differences in the data.
In greedy pruning {\em},
the distant points $A,B$ are evaluated, and we only recurse on the data nearest
the best point. In 
{\em non-greedy pruning}, the whole tree is generated (without
evaluations). $A,B$ evaluation and pruning is
then applied to the largest subtrees with the fewest differences within the most
variable attributes. This is repeated to generate survivors $\sqrt{N}$,
which are then explored with the greedy approach~\cite{lustosa2023optimizing}.\\\hline
\end{tabular}
\caption{Recursive bi-clustering for  model review. Here,
to explore (say)  $N=10,000$ examples, we would only need to pose  \mbox{$2\log_2(N)<30$} questions about
a handful of attributes.}\label{stealth}
\end{table}
% I recommend this
% approach for model review since it can explore
% a  large multi-objective space using just  a few questions 
% (e.g., for $N=10,000$ examples, ask only \mbox{$2\log_2(N)<30$} questions about
% a handful of attributes). 

\
Of course, not all data sets can be explored by a few dozen keys. 
Recently,  we have successfully modeled security violations in 28,750 Mozilla functions with 271 exemplars and 6000 Github commits using just 300 exemplars~\cite{yu2019improving}\footnote{Specifically, after   incremental active
learning, the SVM had   under 300 support vectors.}. Although 300 is not an especially small
number, it is small enough so that, given (say) two analysts and a month, it would be possible
to review them all.

\begin{wrapfigure}{r}{1.25in}
  \includegraphics[width=1.25in]{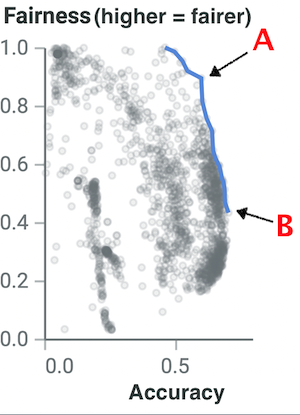}
\caption{Fairness vs. accuracy from 10,000   hyperparameter options~\cite{cruz2021promoting}. ``A'' is desirable but ``B'' is probable.}\label{one}\end{wrapfigure} Note that keys  have obvious implications for software testing. Ling~\cite{ling2023benefits} generates test suites for cyberphysical systems by first finding prototypes. In that work, $N$  candidate  tests are recursively bi-clustered
  to leaves of size $\sqrt{N}$ (using the methods of Table~\ref{stealth}).
Test suites generated from the mode of each cluster are  orders of magnitude faster to generate and just as effective as tests generated
by   more complex (and slower)
methods.

This recursive bi-clustering method has
also been applied to multi-objective
optimization~\cite{agrawal2020better}.   Given
a large enough initial population (e.g. $N=10,000$), recursive bi-clustering is faster and finds better solutions than state-of-the-art genetic algorithms and sequential
model optimization methods~\cite{Chen19,lustosa2023optimizing}
(even though it only evaluates $2\log{10,000}=26$ examples while other methods
might evaluate 100s to 1000s of examples).

Once we have a keys-based multi-objective optimizer, we can offer much support for reviewing models with respect to their fairness.
Figure~\ref{one} comes from Cruz et al.~\cite{cruz2021promoting}.
  That figure 
 shows the effects of 10,000 different   hyperparameter   options applied to five
machine learning algorithms
(random forest; LinReg;
boosted trees; decision trees; feed-forward NN)\footnote{The hyperparameters of     Random Forests,       learners
include
(a)~how many $T$   trees to build (e.g., $T\in \{10,20,40,80,160\}$); (b)~how many features $F$
to use in each tree (e.g., $F \in{2,4,10,20,
\mathit{sqrt}, \mathit{log2}, \mathit{all}}$);
(c)~how to  poll the whole forest (e.g., {\em majority} or {\em weighted majority}); 
(d)~what impurity measures (e.g., {\em gini} or {\em entropy} or {\em log.loss}); (e)~what is the minimum examples needed to branch a sub-tree
(e.g., {\em min}$\in \{2,5,10,20,50,100\}$; (f)~should branches be {\em binary} or {\em n-arty}.
In all, this gives us
$5*7*2*3*6*2 > 2,500$ different ways, just to configure a learner in Figure~\ref{one}. }. 
Adjusting tunings  can change 
learners from   low to   high accuracies and fairness (measured here as the ratio of false
positives between different social groups, such as men and women).   
But with the search methods of the Table~\ref{stealth},
reviewing all those points would require just 20 evaluations-- a number so small that (potentially) it could happen between coffee breaks of a stakeholder review session.

\section{What to Publish at PROMISE v2.0?}

Having presented the methods of the last section,
I rush to add that they are hardly complete.  There is much here that could be better defined/ refined/ improved/ replaced by further research in PROMISE v2.0.

In my view, the goal of a PROMISE v2.0 paper could be ``less is more''; that is, achieve faster, simpler,
better results using some simplification of the existing technique. There are many ways this could be done; e.g. with:
\begin{itemize}
\item Semi-supervised learning methods that let us do much more with much less data~\cite{tu2021frugal,zhu2005semi};
\item Instance or feature selection  to reduce   training data~\cite{olvera2010review,Kohavi97};
\item Distillation methods   reduce model size. For example, a new data set could be synthesized from the branches of a decision tree. By summarizing this new data set, do we find a smaller model? And for alternative neural approaches, 
see~\cite{Shi23,Biswas23}.
\item  {\em Variance studies}
that tests if the improvement of a complex method over a simpler one are   statistically insignificant;
\item Ablation studies~\cite{yedida2023find} to see how
much can be thrown away (of the modeling method, of parts of the training data) while preserving model performance.
\item Studies showing that (say) a 10\% implementation can perform nearly as well as 100\% of an entire system.
\item Some kind of keys-based approach (e.g. see last section).
\end{itemize} 
Human-in-the-loop studies would also be strongly encouraged
in PROMISE v2.0, to test if the smaller models are still acceptable and useful for people. But in line with PROMISE's long
and admirable history of reproducibility, these experiments should
include human surrogates (developed perhaps via data mining)
that can model the strengths and weaknesses of subject matter
experts (and surrogates should be shared as part of the reproducibility package of a paper).

Also encouraged would be a broader range of performance criteria than just (e.g.) accuracy,
Performance should be measured in a multidimensional
manner and include much more than mere predictive performance,
but also runtime, energy usage, discrimination measures, model development cost, etc.

\section{A Counter Argument (More is More)}

My ``less is more'' proposal is  antithetical to much current research in SE and AI.
Data-hungry researchers
in SE assume that ``more is more''; i.e.
if some data are useful, then even more data is even more useful.
For example
``Long-
term JIT models should be trained using a cache of many
changes''~\cite{amasaki2020cross}; and 
``..as long as it is large; the resulting prediction
performance is likely to be boosted more by the size of the
sample'”~\cite{rahman2014comparing}.

A common problem with ``more is more`` is that researchers
often make that assumption without actually testing it.
For example, a recent systematic review~\cite{hou2023large} of the literature on large language models in SE reported 229 research articles from 2017 to 2023. We asked the authors
of that review,  
``how many articles compared their approach to something simpler
non-neural approach?'' and ``in how many of those comparisons
was there any hyperparameter optimization?''. They responded with
a list of 13 papers (\mbox{$\frac{13}{229}\approx 5\%$}) which,
when read, contain some questionable methodological choices. For example,
one of those 13 articles reported that LLMs perform
better than a text mining methods called LDA. But that article used LDA in its ``off-the-shelf'' configuration even though we have seen dramatic improvements in LDA performance with hyperparameter optimization~\cite{agrawal2019dodge,agrawal2018wrong}.

To be clear, I firmly believe that deep learning and generative AI methods
such as ``chain of thought''\footnote{https://github.com/Timothyxxx/Chain-of-ThoughtsPapers} will dramatically change the nature of science (in general) and SE (in particular).  But moving
away from generative tasks to classification, regression, and optimization tasks, my
experimental results strongly suggest that   other non-neural  methods can be just as effective, particularly when combined with hyperparameter optimization. This is an important point since the non-neural methods can yield the succinct symbolic models that humans need to review and understand a model.

But rather than stating all this in an adversarial manner, it might be more useful to ask how ``less is more'' can be beneficial for more
elaborate approaches. Lustosa (work in progress) has explored the hyperparameters of some deep learners using the recursive bi-clustering method described in the Introduction, and found that it was able to configure the deep learners more effectively than other state-of-the-art optimizers, and do so much faster.

\section{A Final Thought}
 
Despite all my   papers on the topic, ``less is more'' is mostly ignored.  Perhaps this is my own fault.
 Initially, I  argued this with simulation studies on artificially created examples, which some people find unconvincing~\cite{menzies2000testing,menzies2004many}. However,   subsequent work reported results from real-world data ~\cite{menziees07strange,menzies2008implications,chen2005finding,menzies2006data,partington2015reduced,jalali2008optimizing,Chen19,lustosa2023optimizing,agrawal2020better,ling2023benefits,yu2019improving,10.1145/3583565}. Furthermore, the latter work included numerous studies that demonstrate that this ``less is more'' approach produces smaller and better models than the current state-of-the-art.

Perhaps there is something deeply embedded in our research culture that encourages and rewards complexity.
Perhaps our
research is driven by the concerns of large software organizations that
prefer complexity (since only those large organizations have the resources to build and maintain complex solutions).

Perhaps also,  in a publication-oriented environment, researchers tend to rush out reports of complex mashups of tools,
rather than refactor and reduce the size of their toolkits. 

Perhaps what we need is a space where we can revisit and reflect on old results, looking for some synthesis that significantly simplifies and improves those results.
To create a journal venue for such papers, I invite submissions to the ``Less is More''   section of the Automated Software Engineering journal\footnote{See \url{https://ause-journal.github.io/simpler.html}}.

And as to an associated conference venue, perhaps that venue   could be PROMISE v2.0?

\section*{Acknowledgements}
Thanks to Hou et al. for quickly responding to a query
on LLMs. Also, thanks to all   the
grad students and coauthors who helped mature, refine, and simplify the ideas
that led to this paper.

\balance
 \bibliographystyle{ACM-Reference-Format}
\bibliography{acmart}


\begin{thebibliography}{62}

%%% the Web address.

\ifx \showCODEN    \undefined \def \showCODEN     #1{\unskip}     \fi
\ifx \showDOI      \undefined \def \showDOI       #1{#1}\fi
\ifx \showISBNx    \undefined \def \showISBNx     #1{\unskip}     \fi
\ifx \showISBNxiii \undefined \def \showISBNxiii  #1{\unskip}     \fi
\ifx \showISSN     \undefined \def \showISSN      #1{\unskip}     \fi
\ifx \showLCCN     \undefined \def \showLCCN      #1{\unskip}     \fi
\ifx \shownote     \undefined \def \shownote      #1{#1}          \fi
\ifx \showarticletitle \undefined \def \showarticletitle #1{#1}   \fi
\ifx \showURL      \undefined \def \showURL       {\relax}        \fi
% invisible to TeX
\providecommand\bibfield[2]{#2}
\providecommand\bibinfo[2]{#2}
\providecommand\natexlab[1]{#1}
\providecommand\showeprint[2][]{arXiv:#2}

\bibitem[Agrawal et~al\mbox{.}(2019)]%
        {agrawal2019dodge}
\bibfield{author}{\bibinfo{person}{Amritanshu Agrawal}, \bibinfo{person}{Wei
  Fu}, \bibinfo{person}{Di Chen}, \bibinfo{person}{Xipeng Shen}, {and}
  \bibinfo{person}{Tim Menzies}.} \bibinfo{year}{2019}\natexlab{}.
\newblock \showarticletitle{How to “dodge” complex software analytics}.
\newblock \bibinfo{journal}{\emph{IEEE TSE}} \bibinfo{volume}{47},
  \bibinfo{number}{10} (\bibinfo{year}{2019}), \bibinfo{pages}{2182--2194}.
\newblock


\bibitem[Agrawal et~al\mbox{.}(2018)]%
        {agrawal2018wrong}
\bibfield{author}{\bibinfo{person}{Amritanshu Agrawal}, \bibinfo{person}{Wei
  Fu}, {and} \bibinfo{person}{Tim Menzies}.} \bibinfo{year}{2018}\natexlab{}.
\newblock \showarticletitle{What is Wrong with Topic Modeling?(and How to Fix
  it Using Search-based Software Engineering)}.
\newblock \bibinfo{journal}{\emph{Information and Software Technology}}
  (\bibinfo{year}{2018}).
\newblock


\bibitem[Agrawal et~al\mbox{.}(2020)]%
        {agrawal2020better}
\bibfield{author}{\bibinfo{person}{Amritanshu Agrawal}, \bibinfo{person}{Tim
  Menzies}, \bibinfo{person}{Leandro~L Minku}, \bibinfo{person}{Markus Wagner},
  {and} \bibinfo{person}{Zhe Yu}.} \bibinfo{year}{2020}\natexlab{}.
\newblock \showarticletitle{Better software analytics via “DUO”: Data
  mining algorithms using/used-by optimizers}.
\newblock \bibinfo{journal}{\emph{Empirical Software Engineering}}
  \bibinfo{volume}{25} (\bibinfo{year}{2020}), \bibinfo{pages}{2099--2136}.
\newblock


\bibitem[Amasaki(2020)]%
        {amasaki2020cross}
\bibfield{author}{\bibinfo{person}{Sousuke Amasaki}.}
  \bibinfo{year}{2020}\natexlab{}.
\newblock \showarticletitle{Cross-version defect prediction: use historical
  data, cross-project data, or both?}
\newblock \bibinfo{journal}{\emph{Empirical Software Engineering}}
  (\bibinfo{year}{2020}), \bibinfo{pages}{1--23}.
\newblock


\bibitem[Biswas and Rajan(2023)]%
        {Biswas23}
\bibfield{author}{\bibinfo{person}{Sumon Biswas} {and} \bibinfo{person}{Hridesh
  Rajan}.} \bibinfo{year}{2023}\natexlab{}.
\newblock \showarticletitle{Fairify: Fairness Verification of Neural Networks}.
  In \bibinfo{booktitle}{\emph{ICSE'23}}. \bibinfo{pages}{1546–1558}.
\newblock
\showISBNx{9781665457019}
\urldef\tempurl%
\url{https://doi.org/10.1109/ICSE48619.2023.00134}
\showDOI{\tempurl}


\bibitem[Boehm(1981)]%
        {Boehm1981}
\bibfield{author}{\bibinfo{person}{B.W. Boehm}.}
  \bibinfo{year}{1981}\natexlab{}.
\newblock \bibinfo{booktitle}{\emph{Software Engineering Economics}}.
\newblock \bibinfo{publisher}{Prentice Hall}.
\newblock


\bibitem[Bowleg(2021)]%
        {doi:10.2105/AJPH.2020.306031}
\bibfield{author}{\bibinfo{person}{Lisa Bowleg}.}
  \bibinfo{year}{2021}\natexlab{}.
\newblock \showarticletitle{Evolving Intersectionality Within Public Health:
  From Analysis to Action}.
\newblock \bibinfo{journal}{\emph{American Journal of Public Health}}
  \bibinfo{volume}{111}, \bibinfo{number}{1} (\bibinfo{year}{2021}),
  \bibinfo{pages}{88--90}.
\newblock
\urldef\tempurl%
\url{https://doi.org/10.2105/AJPH.2020.306031}
\showDOI{\tempurl}
\newblock
\shownote{PMID: 33326269}.


\bibitem[Brighton(2006)]%
        {brighton2006robust}
\bibfield{author}{\bibinfo{person}{Henry Brighton}.}
  \bibinfo{year}{2006}\natexlab{}.
\newblock \showarticletitle{Robust Inference with Simple Cognitive Models.}. In
  \bibinfo{booktitle}{\emph{AAAI spring symposium: Between a rock and a hard
  place: Cognitive science principles meet AI-hard problems}}.
  \bibinfo{pages}{17--22}.
\newblock


\bibitem[Canellas(2021)]%
        {canellas21}
\bibfield{author}{\bibinfo{person}{Marc Canellas}.}
  \bibinfo{year}{2021}\natexlab{}.
\newblock \showarticletitle{Defending IEEE Software Standards in Federal
  Criminal Court}.
\newblock \bibinfo{journal}{\emph{Computer}} \bibinfo{volume}{54},
  \bibinfo{number}{6} (\bibinfo{year}{2021}), \bibinfo{pages}{14--23}.
\newblock
\urldef\tempurl%
\url{https://doi.org/10.1109/MC.2020.3038630}
\showDOI{\tempurl}


\bibitem[Chakraborty et~al\mbox{.}(2021)]%
        {Chakraborty}
\bibfield{author}{\bibinfo{person}{Joymallya Chakraborty},
  \bibinfo{person}{Suvodeep Majumder}, {and} \bibinfo{person}{Tim Menzies}.}
  \bibinfo{year}{2021}\natexlab{}.
\newblock \showarticletitle{Bias in Machine Learning Software: Why? How? What
  to Do?}. In \bibinfo{booktitle}{\emph{FSE'21}}. \bibinfo{address}{New York,
  NY, USA}, \bibinfo{pages}{429–440}.
\newblock
\showISBNx{9781450385626}
\urldef\tempurl%
\url{https://doi.org/10.1145/3468264.3468537}
\showDOI{\tempurl}


\bibitem[Chen et~al\mbox{.}(2019)]%
        {Chen19}
\bibfield{author}{\bibinfo{person}{Jianfeng Chen}, \bibinfo{person}{Vivek
  Nair}, \bibinfo{person}{Rahul Krishna}, {and} \bibinfo{person}{Tim Menzies}.}
  \bibinfo{year}{2019}\natexlab{}.
\newblock \showarticletitle{“Sampling” as a Baseline Optimizer for
  Search-Based Software Engineering}.
\newblock \bibinfo{journal}{\emph{IEEE TSE}} \bibinfo{volume}{45},
  \bibinfo{number}{6} (\bibinfo{year}{2019}), \bibinfo{pages}{597--614}.
\newblock


\bibitem[Chen et~al\mbox{.}(2005)]%
        {chen2005finding}
\bibfield{author}{\bibinfo{person}{Zhihao Chen}, \bibinfo{person}{Tim Menzies},
  \bibinfo{person}{Daniel Port}, {and} \bibinfo{person}{D Boehm}.}
  \bibinfo{year}{2005}\natexlab{}.
\newblock \showarticletitle{Finding the right data for software cost modeling}.
\newblock \bibinfo{journal}{\emph{IEEE software}} \bibinfo{volume}{22},
  \bibinfo{number}{6} (\bibinfo{year}{2005}), \bibinfo{pages}{38--46}.
\newblock


\bibitem[Coaston(2019)]%
        {Coaston19}
\bibfield{author}{\bibinfo{person}{Jane Coaston}.}
  \bibinfo{year}{2019}\natexlab{}.
\newblock
\newblock
\newblock
\shownote{VOX, May 28. On-line at
  \url{https://www.vox.com/the-highlight/2019/5/20/18542843/intersectionality-conservatism-law-race-gender-discrimination}}.


\bibitem[Cruz et~al\mbox{.}(2021)]%
        {cruz2021promoting}
\bibfield{author}{\bibinfo{person}{Andr{\'e}~F Cruz}, \bibinfo{person}{Pedro
  Saleiro}, \bibinfo{person}{Catarina Bel{\'e}m}, \bibinfo{person}{Carlos
  Soares}, {and} \bibinfo{person}{Pedro Bizarro}.}
  \bibinfo{year}{2021}\natexlab{}.
\newblock \showarticletitle{Promoting fairness through hyperparameter
  optimization}. In \bibinfo{booktitle}{\emph{2021 IEEE International
  Conference on Data Mining (ICDM)}}. IEEE, \bibinfo{pages}{1036--1041}.
\newblock


\bibitem[Czerlinski et~al\mbox{.}(1999)]%
        {czerlinski1999good}
\bibfield{author}{\bibinfo{person}{Jean Czerlinski}, \bibinfo{person}{Gerd
  Gigerenzer}, {and} \bibinfo{person}{Daniel~G Goldstein}.}
  \bibinfo{year}{1999}\natexlab{}.
\newblock \showarticletitle{How good are simple heuristics?}. In
  \bibinfo{booktitle}{\emph{Simple Heuristics That Make Us Smart}}.
  \bibinfo{publisher}{Oxford University Press}.
\newblock


\bibitem[Desharnais(1989)]%
        {desharnais1989analyse}
\bibfield{author}{\bibinfo{person}{JM Desharnais}.}
  \bibinfo{year}{1989}\natexlab{}.
\newblock \showarticletitle{Analyse statistique de la productivitie des projets
  informatique a partie de la technique des point des function}.
\newblock \bibinfo{journal}{\emph{Masters Thesis, University of Montreal}}
  (\bibinfo{year}{1989}).
\newblock


\bibitem[Faloutsos and Lin(1995)]%
        {faloutsos1995fastmap}
\bibfield{author}{\bibinfo{person}{Christos Faloutsos} {and}
  \bibinfo{person}{King-Ip Lin}.} \bibinfo{year}{1995}\natexlab{}.
\newblock \showarticletitle{FastMap: A fast algorithm for indexing, data-mining
  and visualization of traditional and multimedia datasets}. In
  \bibinfo{booktitle}{\emph{Proceedings of the 1995 ACM SIGMOD international
  conference on Management of data}}. \bibinfo{pages}{163--174}.
\newblock


\bibitem[Gay et~al\mbox{.}(2010)]%
        {gay2010automatically}
\bibfield{author}{\bibinfo{person}{Gregory Gay}, \bibinfo{person}{Tim Menzies},
  \bibinfo{person}{Misty Davies}, {and} \bibinfo{person}{Karen Gundy-Burlet}.}
  \bibinfo{year}{2010}\natexlab{}.
\newblock \showarticletitle{Automatically finding the control variables for
  complex system behavior}.
\newblock \bibinfo{journal}{\emph{Automated Software Engineering}}
  \bibinfo{volume}{17} (\bibinfo{year}{2010}), \bibinfo{pages}{439--468}.
\newblock


\bibitem[Gigerenzer(2008)]%
        {gigerenzer2008heuristics}
\bibfield{author}{\bibinfo{person}{Gerd Gigerenzer}.}
  \bibinfo{year}{2008}\natexlab{}.
\newblock \showarticletitle{Why heuristics work}.
\newblock \bibinfo{journal}{\emph{Perspectives on psychological science}}
  \bibinfo{volume}{3}, \bibinfo{number}{1} (\bibinfo{year}{2008}),
  \bibinfo{pages}{20--29}.
\newblock


\bibitem[Gigerenzer et~al\mbox{.}(1999)]%
        {gigerenzer1999good}
\bibfield{author}{\bibinfo{person}{Gerd Gigerenzer}, \bibinfo{person}{Jean
  Czerlinski}, {and} \bibinfo{person}{Laura Martignon}.}
  \bibinfo{year}{1999}\natexlab{}.
\newblock \showarticletitle{How good are fast and frugal heuristics}.
\newblock \bibinfo{journal}{\emph{Decision science and technology: Reflections
  on the contributions of Ward Edwards}} (\bibinfo{year}{1999}),
  \bibinfo{pages}{81--103}.
\newblock


\bibitem[Gigerenzer and Gaissmaier(2011)]%
        {gigerenzer2011heuristic}
\bibfield{author}{\bibinfo{person}{Gerd Gigerenzer} {and}
  \bibinfo{person}{Wolfgang Gaissmaier}.} \bibinfo{year}{2011}\natexlab{}.
\newblock \showarticletitle{Heuristic decision making}.
\newblock \bibinfo{journal}{\emph{Annual review of psychology}}
  \bibinfo{volume}{62} (\bibinfo{year}{2011}), \bibinfo{pages}{451--482}.
\newblock


\bibitem[Green(2022)]%
        {green2022flaws}
\bibfield{author}{\bibinfo{person}{Ben Green}.}
  \bibinfo{year}{2022}\natexlab{}.
\newblock \showarticletitle{The flaws of policies requiring human oversight of
  government algorithms}.
\newblock \bibinfo{journal}{\emph{Computer Law \& Security Review}}
  \bibinfo{volume}{45} (\bibinfo{year}{2022}), \bibinfo{pages}{105681}.
\newblock


\bibitem[Holte(1993)]%
        {Holte1993VerySC}
\bibfield{author}{\bibinfo{person}{Robert~C. Holte}.}
  \bibinfo{year}{1993}\natexlab{}.
\newblock \showarticletitle{Very Simple Classification Rules Perform Well on
  Most Commonly Used Datasets}.
\newblock \bibinfo{journal}{\emph{Machine Learning}}  \bibinfo{volume}{11}
  (\bibinfo{year}{1993}), \bibinfo{pages}{63--90}.
\newblock
\urldef\tempurl%
\url{https://api.semanticscholar.org/CorpusID:6596}
\showURL{%
\tempurl}


\bibitem[Hou et~al\mbox{.}(2023)]%
        {hou2023large}
\bibfield{author}{\bibinfo{person}{Xinyi Hou}, \bibinfo{person}{Yanjie Zhao},
  \bibinfo{person}{Yue Liu}, \bibinfo{person}{Zhou Yang},
  \bibinfo{person}{Kailong Wang}, \bibinfo{person}{Li Li},
  \bibinfo{person}{Xiapu Luo}, \bibinfo{person}{David Lo},
  \bibinfo{person}{John Grundy}, {and} \bibinfo{person}{Haoyu Wang}.}
  \bibinfo{year}{2023}\natexlab{}.
\newblock \bibinfo{title}{Large Language Models for Software Engineering: A
  Systematic Literature Review}.
\newblock
\newblock
\showeprint[arxiv]{2308.10620}~[cs.SE]


\bibitem[Jalali et~al\mbox{.}(2008)]%
        {jalali2008optimizing}
\bibfield{author}{\bibinfo{person}{Omid Jalali}, \bibinfo{person}{Tim Menzies},
  {and} \bibinfo{person}{Martin Feather}.} \bibinfo{year}{2008}\natexlab{}.
\newblock \showarticletitle{Optimizing requirements decisions with keys}. In
  \bibinfo{booktitle}{\emph{Proceedings of the 4th international workshop on
  Predictor models in software engineering}}. \bibinfo{pages}{79--86}.
\newblock


\bibitem[Johnson and Menzies(2023)]%
        {mathews23}
\bibfield{author}{\bibinfo{person}{Brittany Johnson} {and} \bibinfo{person}{Tim
  Menzies}.} \bibinfo{year}{2023}\natexlab{}.
\newblock \showarticletitle{Unfairness is everywhere. So what to do? An
  interview with Jeanna Matthews}.
\newblock \bibinfo{journal}{\emph{IEEE Software}} (\bibinfo{year}{2023}).
\newblock
Issue Nov/Dev.


\bibitem[Jureczko and Madeyski(2010)]%
        {jureczko2010towards}
\bibfield{author}{\bibinfo{person}{Marian Jureczko} {and} \bibinfo{person}{Lech
  Madeyski}.} \bibinfo{year}{2010}\natexlab{}.
\newblock \showarticletitle{Towards identifying software project clusters with
  regard to defect prediction}. In \bibinfo{booktitle}{\emph{Proceedings of the
  6th international conference on predictive models in software engineering}}.
  \bibinfo{pages}{1--10}.
\newblock


\bibitem[Kohavi and John(1997)]%
        {Kohavi97}
\bibfield{author}{\bibinfo{person}{Ron Kohavi} {and} \bibinfo{person}{George~H.
  John}.} \bibinfo{year}{1997}\natexlab{}.
\newblock \showarticletitle{Wrappers for Feature Subset Selection}.
\newblock \bibinfo{journal}{\emph{Artificial Intelligence}}
  \bibinfo{volume}{97}, \bibinfo{number}{1-2} (\bibinfo{year}{1997}),
  \bibinfo{pages}{273--324}.
\newblock


\bibitem[Larkin et~al\mbox{.}(1980)]%
        {Larkin1335}
\bibfield{author}{\bibinfo{person}{Jill Larkin}, \bibinfo{person}{John
  McDermott}, \bibinfo{person}{Dorothea~P. Simon}, {and}
  \bibinfo{person}{Herbert~A. Simon}.} \bibinfo{year}{1980}\natexlab{}.
\newblock \showarticletitle{Expert and Novice Performance in Solving Physics
  Problems}.
\newblock \bibinfo{journal}{\emph{Science}} \bibinfo{volume}{208},
  \bibinfo{number}{4450} (\bibinfo{year}{1980}), \bibinfo{pages}{1335--1342}.
\newblock
\urldef\tempurl%
\url{https://doi.org/10.1126/science.208.4450.1335}
\showDOI{\tempurl}
\showeprint{http://science.sciencemag.org/content/208/4450/1335.full.pdf}


\bibitem[Ling and Menzies(2023)]%
        {ling2023benefits}
\bibfield{author}{\bibinfo{person}{Xiao Ling} {and} \bibinfo{person}{Tim
  Menzies}.} \bibinfo{year}{2023}\natexlab{}.
\newblock \bibinfo{title}{On the Benefits of Semi-Supervised Test Case
  Generation for Cyber-Physical Systems}.
\newblock
\newblock
\showeprint[arxiv]{2305.03714}~[cs.SE]


\bibitem[Lustosa and Menzies(2023)]%
        {lustosa2023optimizing}
\bibfield{author}{\bibinfo{person}{Andre Lustosa} {and} \bibinfo{person}{Tim
  Menzies}.} \bibinfo{year}{2023}\natexlab{}.
\newblock \showarticletitle{Optimizing Predictions for Very Small Data Sets: a
  case study on Open-Source Project Health Prediction}.
\newblock \bibinfo{journal}{\emph{arXiv preprint arXiv:2301.06577}}
  (\bibinfo{year}{2023}).
\newblock


\bibitem[Ma et~al\mbox{.}(2014)]%
        {ma2014changing}
\bibfield{author}{\bibinfo{person}{Wei~Ji Ma}, \bibinfo{person}{Masud Husain},
  {and} \bibinfo{person}{Paul~M Bays}.} \bibinfo{year}{2014}\natexlab{}.
\newblock \showarticletitle{Changing concepts of working memory}.
\newblock \bibinfo{journal}{\emph{Nature neuroscience}} \bibinfo{volume}{17},
  \bibinfo{number}{3} (\bibinfo{year}{2014}), \bibinfo{pages}{347--356}.
\newblock


\bibitem[Majumder et~al\mbox{.}(2023)]%
        {10.1145/3585006}
\bibfield{author}{\bibinfo{person}{Suvodeep Majumder},
  \bibinfo{person}{Joymallya Chakraborty}, \bibinfo{person}{Gina~R. Bai},
  \bibinfo{person}{Kathryn~T. Stolee}, {and} \bibinfo{person}{Tim Menzies}.}
  \bibinfo{year}{2023}\natexlab{}.
\newblock \showarticletitle{Fair Enough: Searching for Sufficient Measures of
  Fairness}.
\newblock \bibinfo{journal}{\emph{ACM Trans. Softw. Eng. Methodol.}}
  (\bibinfo{date}{mar} \bibinfo{year}{2023}).
\newblock
\showISSN{1049-331X}
\urldef\tempurl%
\url{https://doi.org/10.1145/3585006}
\showDOI{\tempurl}
\newblock
\shownote{Just Accepted}.


\bibitem[Martignon et~al\mbox{.}(2008)]%
        {martignon2008categorization}
\bibfield{author}{\bibinfo{person}{Laura Martignon},
  \bibinfo{person}{Konstantinos~V Katsikopoulos}, {and} \bibinfo{person}{Jan~K
  Woike}.} \bibinfo{year}{2008}\natexlab{}.
\newblock \showarticletitle{Categorization with limited resources: A family of
  simple heuristics}.
\newblock \bibinfo{journal}{\emph{Journal of Mathematical Psychology}}
  \bibinfo{volume}{52}, \bibinfo{number}{6} (\bibinfo{year}{2008}),
  \bibinfo{pages}{352--361}.
\newblock


\bibitem[Martignon et~al\mbox{.}(2003)]%
        {martignon2003naive}
\bibfield{author}{\bibinfo{person}{Laura Martignon}, \bibinfo{person}{Oliver
  Vitouch}, \bibinfo{person}{Masanori Takezawa}, {and}
  \bibinfo{person}{Malcolm~R Forster}.} \bibinfo{year}{2003}\natexlab{}.
\newblock \showarticletitle{Naive and yet enlightened: From natural frequencies
  to fast and frugal decision trees}.
\newblock \bibinfo{journal}{\emph{Thinking: Psychological perspective on
  reasoning, judgment, and decision making}} (\bibinfo{year}{2003}),
  \bibinfo{pages}{189--211}.
\newblock


\bibitem[Matthews(2023)]%
        {matthewsshould}
\bibfield{author}{\bibinfo{person}{Jeanna Matthews}.}
  \bibinfo{year}{2023}\natexlab{}.
\newblock \showarticletitle{How should we regulate AI? Practical Strategies for
  Regulation and Risk Management from the IEEE1012 Standard for System,
  Software, and Hardware Verification and Validation.}
\newblock  (\bibinfo{year}{2023}).
\newblock
\newblock
\shownote{\url{https://ieeeusa.org/assets/public-policy/committees/aipc/How-Should-We-Regulate-AI.pdf}}.


\bibitem[Menzies et~al\mbox{.}(2000)]%
        {menzies2000testing}
\bibfield{author}{\bibinfo{person}{Tim Menzies}, \bibinfo{person}{Bojan Cukic},
  \bibinfo{person}{Harshinder Singh}, {and} \bibinfo{person}{John Powell}.}
  \bibinfo{year}{2000}\natexlab{}.
\newblock \showarticletitle{Testing nondeterminate systems}. In
  \bibinfo{booktitle}{\emph{Proceedings 11th International Symposium on
  Software Reliability Engineering. ISSRE 2000}}. IEEE,
  \bibinfo{pages}{222--231}.
\newblock


\bibitem[Menzies and Di~Stefano(2004)]%
        {menzies2004good}
\bibfield{author}{\bibinfo{person}{Tim Menzies} {and} \bibinfo{person}{Justin~S
  Di~Stefano}.} \bibinfo{year}{2004}\natexlab{}.
\newblock \showarticletitle{How good is your blind spot sampling policy}. In
  \bibinfo{booktitle}{\emph{Eighth IEEE International Symposium on High
  Assurance Systems Engineering, 2004. Proceedings.}} IEEE,
  \bibinfo{pages}{129--138}.
\newblock


\bibitem[Menzies et~al\mbox{.}(2006)]%
        {menzies2006data}
\bibfield{author}{\bibinfo{person}{Tim Menzies}, \bibinfo{person}{Jeremy
  Greenwald}, {and} \bibinfo{person}{Art Frank}.}
  \bibinfo{year}{2006}\natexlab{}.
\newblock \showarticletitle{Data mining static code attributes to learn defect
  predictors}.
\newblock \bibinfo{journal}{\emph{IEEE transactions on software engineering}}
  \bibinfo{volume}{33}, \bibinfo{number}{1} (\bibinfo{year}{2006}),
  \bibinfo{pages}{2--13}.
\newblock


\bibitem[Menzies et~al\mbox{.}(2007)]%
        {menziees07strange}
\bibfield{author}{\bibinfo{person}{Tim Menzies}, \bibinfo{person}{David Owen},
  {and} \bibinfo{person}{Julian Richardson}.} \bibinfo{year}{2007}\natexlab{}.
\newblock \showarticletitle{The Strangest Thing About Software}.
\newblock \bibinfo{journal}{\emph{Computer}} \bibinfo{volume}{40},
  \bibinfo{number}{1} (\bibinfo{year}{2007}), \bibinfo{pages}{54--60}.
\newblock
\urldef\tempurl%
\url{https://doi.org/10.1109/MC.2007.37}
\showDOI{\tempurl}


\bibitem[Menzies and Singh(2004)]%
        {menzies2004many}
\bibfield{author}{\bibinfo{person}{Tim Menzies} {and}
  \bibinfo{person}{Harhsinder Singh}.} \bibinfo{year}{2004}\natexlab{}.
\newblock \showarticletitle{Many maybes mean (mostly) the same thing}.
\newblock \bibinfo{journal}{\emph{Soft Computing in Software Engineering}}
  (\bibinfo{year}{2004}), \bibinfo{pages}{125--150}.
\newblock


\bibitem[Menzies et~al\mbox{.}(2008)]%
        {menzies2008implications}
\bibfield{author}{\bibinfo{person}{Tim Menzies}, \bibinfo{person}{Burak
  Turhan}, \bibinfo{person}{Ayse Bener}, \bibinfo{person}{Gregory Gay},
  \bibinfo{person}{Bojan Cukic}, {and} \bibinfo{person}{Yue Jiang}.}
  \bibinfo{year}{2008}\natexlab{}.
\newblock \showarticletitle{Implications of ceiling effects in defect
  predictors}. In \bibinfo{booktitle}{\emph{Proceedings of the 4th
  international workshop on Predictor models in software engineering}}.
  \bibinfo{pages}{47--54}.
\newblock


\bibitem[N.C. and Menzies(2023)]%
        {10.1145/3583565}
\bibfield{author}{\bibinfo{person}{Shrikanth N.C.} {and} \bibinfo{person}{Tim
  Menzies}.} \bibinfo{year}{2023}\natexlab{}.
\newblock \showarticletitle{Assessing the Early Bird Heuristic (for Predicting
  Project Quality)}.
\newblock \bibinfo{journal}{\emph{ACM Trans. Softw. Eng. Methodol.}}
  \bibinfo{volume}{32}, \bibinfo{number}{5}, Article \bibinfo{articleno}{116}
  (\bibinfo{date}{jul} \bibinfo{year}{2023}), \bibinfo{numpages}{39}~pages.
\newblock
\showISSN{1049-331X}
\urldef\tempurl%
\url{https://doi.org/10.1145/3583565}
\showDOI{\tempurl}


\bibitem[Neth and Gigerenzer(2015)]%
        {neth2015heuristics}
\bibfield{author}{\bibinfo{person}{Hansj{\"o}rg Neth} {and}
  \bibinfo{person}{Gerd Gigerenzer}.} \bibinfo{year}{2015}\natexlab{}.
\newblock \showarticletitle{Heuristics: Tools for an uncertain world}.
\newblock \bibinfo{journal}{\emph{Emerging trends in the social and behavioral
  sciences: An interdisciplinary, searchable, and linkable resource}}
  (\bibinfo{year}{2015}).
\newblock


\bibitem[Noble(2018)]%
        {noble2018algorithms}
\bibfield{author}{\bibinfo{person}{Safiya~Umoja Noble}.}
  \bibinfo{year}{2018}\natexlab{}.
\newblock \bibinfo{booktitle}{\emph{Algorithms of oppression. How search
  engines reinforce racism}}.
\newblock \bibinfo{publisher}{New York University Press}, \bibinfo{address}{New
  York}.
\newblock
\showISBNx{978-1-4798-4994-9}
\urldef\tempurl%
\url{http://algorithmsofoppression.com/}
\showURL{%
\tempurl}


\bibitem[Olvera-L{\'o}pez et~al\mbox{.}(2010)]%
        {olvera2010review}
\bibfield{author}{\bibinfo{person}{J~Arturo Olvera-L{\'o}pez},
  \bibinfo{person}{J~Ariel Carrasco-Ochoa}, \bibinfo{person}{J~Francisco
  Mart{\'\i}nez-Trinidad}, {and} \bibinfo{person}{Josef Kittler}.}
  \bibinfo{year}{2010}\natexlab{}.
\newblock \showarticletitle{A review of instance selection methods}.
\newblock \bibinfo{journal}{\emph{Artificial Intelligence Review}}
  \bibinfo{volume}{34} (\bibinfo{year}{2010}), \bibinfo{pages}{133--143}.
\newblock


\bibitem[Papakroni(2013)]%
        {papakroni2013data}
\bibfield{author}{\bibinfo{person}{Vasil Papakroni}.}
  \bibinfo{year}{2013}\natexlab{}.
\newblock \bibinfo{booktitle}{\emph{Data carving: Identifying and removing
  irrelevancies in the data}}.
\newblock \bibinfo{publisher}{West Virginia University}.
\newblock


\bibitem[Partington et~al\mbox{.}(2015)]%
        {partington2015reduced}
\bibfield{author}{\bibinfo{person}{Susan~N Partington}, \bibinfo{person}{Tim~J
  Menzies}, \bibinfo{person}{Trina~A Colburn}, \bibinfo{person}{Brian~E
  Saelens}, {and} \bibinfo{person}{Karen Glanz}.}
  \bibinfo{year}{2015}\natexlab{}.
\newblock \showarticletitle{Reduced-item food audits based on the nutrition
  environment measures surveys}.
\newblock \bibinfo{journal}{\emph{American Journal of Preventive Medicine}}
  \bibinfo{volume}{49}, \bibinfo{number}{4} (\bibinfo{year}{2015}),
  \bibinfo{pages}{e23--e33}.
\newblock


\bibitem[Phillips et~al\mbox{.}(2017)]%
        {phillips2017FFTrees}
\bibfield{author}{\bibinfo{person}{Nathaniel~D Phillips},
  \bibinfo{person}{Hansjoerg Neth}, \bibinfo{person}{Jan~K Woike}, {and}
  \bibinfo{person}{Wolfgang Gaissmaier}.} \bibinfo{year}{2017}\natexlab{}.
\newblock \showarticletitle{{FFTrees}: A toolbox to create, visualize, and
  evaluate fast-and-frugal decision trees}.
\newblock \bibinfo{journal}{\emph{Judgment and Decision Making}}
  \bibinfo{volume}{12}, \bibinfo{number}{4} (\bibinfo{year}{2017}),
  \bibinfo{pages}{344--368}.
\newblock


\bibitem[Platt(2005)]%
        {platt2005fastmap}
\bibfield{author}{\bibinfo{person}{John Platt}.}
  \bibinfo{year}{2005}\natexlab{}.
\newblock \showarticletitle{Fastmap, metricmap, and landmark mds are all
  nystr{\"o}m algorithms}. In \bibinfo{booktitle}{\emph{International Workshop
  on Artificial Intelligence and Statistics}}. PMLR, \bibinfo{pages}{261--268}.
\newblock


\bibitem[Rahman et~al\mbox{.}(2014)]%
        {rahman2014comparing}
\bibfield{author}{\bibinfo{person}{Foyzur Rahman}, \bibinfo{person}{Sameer
  Khatri}, \bibinfo{person}{Earl~T Barr}, {and} \bibinfo{person}{Premkumar
  Devanbu}.} \bibinfo{year}{2014}\natexlab{}.
\newblock \showarticletitle{Comparing static bug finders and statistical
  prediction}. In \bibinfo{booktitle}{\emph{ICSE'14}}.
\newblock


\bibitem[Rees-Jones et~al\mbox{.}(2017)]%
        {rees2017better}
\bibfield{author}{\bibinfo{person}{Mitch Rees-Jones}, \bibinfo{person}{Matthew
  Martin}, {and} \bibinfo{person}{Tim Menzies}.}
  \bibinfo{year}{2017}\natexlab{}.
\newblock \showarticletitle{Better predictors for issue lifetime}.
\newblock \bibinfo{journal}{\emph{arXiv preprint arXiv:1702.07735}}
  (\bibinfo{year}{2017}).
\newblock


\bibitem[Rudin(2019)]%
        {rudin2019explaining}
\bibfield{author}{\bibinfo{person}{Cynthia Rudin}.}
  \bibinfo{year}{2019}\natexlab{}.
\newblock \showarticletitle{Stop explaining black box machine learning models
  for high stakes decisions and use interpretable models instead}.
\newblock \bibinfo{journal}{\emph{Nature Machine Intelligence}}
  \bibinfo{volume}{1}, \bibinfo{number}{5} (\bibinfo{year}{2019}),
  \bibinfo{pages}{206--215}.
\newblock
\showISSN{25225839}
\urldef\tempurl%
\url{https://doi.org/10.1038/s42256-019-0048-x}
\showDOI{\tempurl}


\bibitem[Shi et~al\mbox{.}(2023)]%
        {Shi23}
\bibfield{author}{\bibinfo{person}{Jieke Shi}, \bibinfo{person}{Zhou Yang},
  \bibinfo{person}{Bowen Xu}, \bibinfo{person}{Hong~Jin Kang}, {and}
  \bibinfo{person}{David Lo}.} \bibinfo{year}{2023}\natexlab{}.
\newblock \showarticletitle{Compressing Pre-Trained Models of Code into 3 MB}.
  In \bibinfo{booktitle}{\emph{ASE'22}}. Article \bibinfo{articleno}{24},
  \bibinfo{numpages}{12}~pages.
\newblock
\showISBNx{9781450394758}
\urldef\tempurl%
\url{https://doi.org/10.1145/3551349.3556964}
\showDOI{\tempurl}


\bibitem[Simon(1956)]%
        {simon1956rational}
\bibfield{author}{\bibinfo{person}{Herbert~A. Simon}.}
  \bibinfo{year}{1956}\natexlab{}.
\newblock \showarticletitle{Rational choice and the structure of the
  environment}.
\newblock \bibinfo{journal}{\emph{Psychological Review}} \bibinfo{volume}{63},
  \bibinfo{number}{2} (\bibinfo{year}{1956}), \bibinfo{pages}{129--138}.
\newblock


\bibitem[Tawosi et~al\mbox{.}(2023)]%
        {Tawosi23}
\bibfield{author}{\bibinfo{person}{Vali Tawosi}, \bibinfo{person}{Rebecca
  Moussa}, {and} \bibinfo{person}{Federica Sarro}.}
  \bibinfo{year}{2023}\natexlab{}.
\newblock \showarticletitle{Agile Effort Estimation: Have We Solved the Problem
  Yet? Insights From a Replication Study}.
\newblock \bibinfo{journal}{\emph{IEEE Transactions on Software Engineering}}
  \bibinfo{volume}{49}, \bibinfo{number}{4} (\bibinfo{year}{2023}),
  \bibinfo{pages}{2677--2697}.
\newblock
\urldef\tempurl%
\url{https://doi.org/10.1109/TSE.2022.3228739}
\showDOI{\tempurl}


\bibitem[Tu and Menzies(2021)]%
        {tu2021frugal}
\bibfield{author}{\bibinfo{person}{Huy Tu} {and} \bibinfo{person}{Tim
  Menzies}.} \bibinfo{year}{2021}\natexlab{}.
\newblock \showarticletitle{FRUGAL: unlocking semi-supervised learning for
  software analytics}. In \bibinfo{booktitle}{\emph{2021 36th IEEE/ACM
  International Conference on Automated Software Engineering (ASE)}}. IEEE,
  \bibinfo{pages}{394--406}.
\newblock


\bibitem[Xia et~al\mbox{.}(2022)]%
        {xia22}
\bibfield{author}{\bibinfo{person}{Tianpei Xia}, \bibinfo{person}{Rui Shu},
  \bibinfo{person}{Xipeng Shen}, {and} \bibinfo{person}{Tim Menzies}.}
  \bibinfo{year}{2022}\natexlab{}.
\newblock \showarticletitle{Sequential Model Optimization for Software Effort
  Estimation}.
\newblock \bibinfo{journal}{\emph{IEEE Transactions on Software Engineering}}
  \bibinfo{volume}{48}, \bibinfo{number}{6} (\bibinfo{year}{2022}),
  \bibinfo{pages}{1994--2009}.
\newblock
\urldef\tempurl%
\url{https://doi.org/10.1109/TSE.2020.3047072}
\showDOI{\tempurl}


\bibitem[Xu et~al\mbox{.}(2021)]%
        {Xu21}
\bibfield{author}{\bibinfo{person}{Zhou Xu}, \bibinfo{person}{Li Li},
  \bibinfo{person}{Meng Yan}, \bibinfo{person}{Jin Liu}, \bibinfo{person}{Xiapu
  Luo}, \bibinfo{person}{John Grundy}, \bibinfo{person}{Yifeng Zhang}, {and}
  \bibinfo{person}{Xiaohong Zhang}.} \bibinfo{year}{2021}\natexlab{}.
\newblock \showarticletitle{A comprehensive comparative study of
  clustering-based unsupervised defect prediction models}.
\newblock \bibinfo{journal}{\emph{Journal of Systems and Software}}
  \bibinfo{volume}{172} (\bibinfo{year}{2021}), \bibinfo{pages}{110862}.
\newblock
\showISSN{0164-1212}
\urldef\tempurl%
\url{https://doi.org/10.1016/j.jss.2020.110862}
\showDOI{\tempurl}


\bibitem[Yedida et~al\mbox{.}(2023)]%
        {yedida2023find}
\bibfield{author}{\bibinfo{person}{Rahul Yedida}, \bibinfo{person}{Hong~Jin
  Kang}, \bibinfo{person}{Huy Tu}, \bibinfo{person}{Xueqi Yang},
  \bibinfo{person}{David Lo}, {and} \bibinfo{person}{Tim Menzies}.}
  \bibinfo{year}{2023}\natexlab{}.
\newblock \showarticletitle{How to find actionable static analysis warnings: A
  case study with FindBugs}.
\newblock \bibinfo{journal}{\emph{IEEE Transactions on Software Engineering}}
  (\bibinfo{year}{2023}).
\newblock


\bibitem[Yu et~al\mbox{.}(2019)]%
        {yu2019improving}
\bibfield{author}{\bibinfo{person}{Zhe Yu}, \bibinfo{person}{Christopher
  Theisen}, \bibinfo{person}{Laurie Williams}, {and} \bibinfo{person}{Tim
  Menzies}.} \bibinfo{year}{2019}\natexlab{}.
\newblock \showarticletitle{Improving vulnerability inspection efficiency using
  active learning}.
\newblock \bibinfo{journal}{\emph{IEEE Transactions on Software Engineering}}
  \bibinfo{volume}{47}, \bibinfo{number}{11} (\bibinfo{year}{2019}),
  \bibinfo{pages}{2401--2420}.
\newblock


\bibitem[Zhu(2005)]%
        {zhu2005semi}
\bibfield{author}{\bibinfo{person}{Xiaojin~Jerry Zhu}.}
  \bibinfo{year}{2005}\natexlab{}.
\newblock \showarticletitle{Semi-supervised learning literature survey}.
\newblock  (\bibinfo{year}{2005}).
\newblock


\end{thebibliography}

\end{document}